\documentclass[aps,showpacs,twocolumn]{revtex4}
\usepackage{amssymb}
\usepackage{natbib}
\usepackage{amsmath}
\usepackage{amsfonts}
\usepackage{graphicx}
\usepackage{mathrsfs}
\usepackage{dcolumn}
\usepackage{bm}
\usepackage{color}
\definecolor{rot}{rgb}{0.75,0.05,0.25}
\definecolor{hellgrau}{gray}{0.5}
\definecolor{blau}{rgb}{0,0,0.7}

\definecolor{rot}{rgb}{0.75,0.05,0.25}
\definecolor{hellgrau}{gray}{0.5}
\definecolor{blau}{rgb}{0,0,0.7}

\newtheorem{theo}{Theorem}

\begin{document}

\title{Finite Bath Fluctuation Theorem}
\author{Michele Campisi}\email{Michele.Campisi@physik.uni-augsburg.de}
\author{Peter Talkner}\author{Peter H\"anggi}
\address{Institute of Physics, University of Augsburg, Universit\"atsstrasse 1, D-86153 Augsburg, Germany}
\date{\today }

\begin{abstract}
We demonstrate that a Finite Bath Fluctuation Theorem  of the Crooks type holds for systems that have been thermalized via weakly coupling it to  a  bath with energy independent finite specific heat. We show that this theorem reduces to the known canonical and microcanonical fluctuation theorems in the two respective limiting cases of infinite and vanishing specific heat of the bath.
The  result is elucidated by applying it to a 2D hard disk colliding elastically with few other hard disks in a rectangular box with perfectly reflecting walls.
\end{abstract}

\maketitle

\section{Introduction}
During the last decade a number of fluctuation theorems have been reported in the literature, which have contributed a good deal to a better understanding of nonequilibrium thermodynamics \cite{Evans:1993df,Gallavotti:1995hs,Jarzynski1997,Jarzynski2007,Crooks:1999fq, talkner2007,TalknerLutzHanggiPRE2007,CampisiTalknerHanggiPRL09}. These can be roughly divided in two categories: steady state fluctuation theorems and transient fluctuation theorems. The former apply to systems in nonequilibrium steady states and give information on the system fluctuations in the asymptotic regime of very large times (see \cite{MaesSEMPOINCARE03,MarconiPHYSREP08,JarzynskiEPJB2008} for reviews on this topic). The latter apply to systems that are temporarily driven out of equilibrium and give information about the fluctuations of work generated by the driving forces. The most representative example of the latter kind is the Crooks fluctuation theorem \cite{Crooks:1999fq, talkner2007}, that applies to systems that are initially in a canonical state. Although the canonical case is by far the most common case, one may need to study situations where the system is initially distributed according to some other statistics, instead. For example the system might be initially a microcanonical state of well defined energy. In this latter case it has been shown that a microcanonical fluctuation theorem of the type of Crooks also exists \cite{Cleuren:2006le,Talkner:2008gb}. One naturally then wonders whether the same type of transient fluctuation theorem exists as well for yet other types of statistics.

In this work we focus on the probability distribution function (pdf) that describes the statistics of a subsystem of a total classical \emph{ergodic} system with fixed energy. For the case where the interaction between the subsystem and the rest of the total system (which we will refer to as the \emph{bath}) is weak, and the specific heat of the bath is independent of the energy (as for an ideal gas, or for a bath composed of hard spheres), the derivation of the pdf is a standard problem of statistical mechanics \cite{thompson,Prosper93}.
We make no assumptions regarding the size of the bath; in particular we do not assume that it is much larger or smaller than that of the system of interest as assumed in the canonical and microcanonical cases respectively. For this reason we refer to this type of bath as to a \emph{finite heat bath}, and to the statistics of the subsystem as to the \emph{finite bath statistics} (see Eq. (\ref{eq:rho-c}) below). For this statistics we show that a fluctuation theorem of the type of Crooks, i.e., a \emph{Finite Bath Fluctuation Theorem} holds. This Finite Bath Fluctuation Theorem includes the Crooks canonical fluctuation theorem and the microcanonical fluctuation theorem, as the two limiting cases in which the bath specific heat goes to infinity and zero, respectively.

The present work is organized as follows. In Sec. \ref{sec:FHBS} we review the derivation of finite heat bath statistics and recall some of its properties. In Sec. \ref{sec:FT} we derive the corresponding Finite Bath Fluctuation Theorem, and show that it reduces to microcanonical and canonical fluctuation theorems in the limits of vanishing and infinite baths, respectively. In Sec. \ref{sec:example} we apply the theory to a specific example (i.e., a 2D hard disk elastically colliding with few other hard disks in a box) and test the validity of the Finite Bath Fluctuation Theorem, both analytically and numerically. Sec. \ref{sec:discussion} contains a discussion of the obtained results. The conclusions are drawn in Sec. \ref{sec:conl}.

\section{\label{sec:FHBS}Finite Bath Statistics}
Let us consider a finite classical Hamiltonian system of total $N_{tot}$ particles and  total energy $E_{tot}$ composed of two \emph{weakly interacting} subsystems: the ``system of interest'' (or simply the system) and the ``bath''. Assuming that the total system is ergodic, the probability density function (pdf) of the system is given  in terms of density of states, $\Omega_B(E)$, of the bath and density of states of the total system, $\Omega_{tot}(E)$, as \cite{Khinchin}:
\begin{equation}
\rho(\textbf{z},\lambda)=\frac{\Omega_B (E_{tot}-H({ \textbf{z}},\lambda))}{\Omega_{tot}(E_{tot})}
\label{eq:rho1}
\end{equation}
where $\textbf{z}=(p_1, ... , p_s, q_1, ... ,q_s)$ stands for the $2s$ dimensional phase space point of the system. Here we assume that the (sub)-system Hamiltonian $H({\textbf{z}},\lambda)$ may depend on some externally controllable parameter $\lambda$ (this could be for instance the volume of a vessel that contains the system, or an applied magnetic or electric field). For example, in the case of a bath composed of $n$ hard spheres in $3$ dimensions, it is $\Omega_B(E_B) \propto E_B^{3n/2-1}$ (see Appendix \ref{app:A}), and one finds from (\ref{eq:rho1}) \cite{thompson}:
\begin{equation}
\rho(\textbf{z},\lambda)= \frac{[E_{tot}-H({ \textbf{z}},\lambda) ]^{3n/2-1}_+}{\int d  \textbf{z} [E_{tot}-H({ \textbf{z}},\lambda) ]^{3n/2-1}_+ }
\label{eq:rho-special-case}
\end{equation}
which is a known result of classical statistical mechanics \cite{note1}.
The symbol $[x]_+$ is defined as $[x]_+:=x\theta(x)$, with $\theta(x)$ denoting Heaviside step function. Note that, in this case, the specific heat of the bath $C(E_B)$, is energy-independent and equal to $3n/2$ \cite{note2}.
More generally one has the following theorem \cite{Almeida01,Campisi06.2}:
\begin{theo}
The system pdf is given by
\begin{equation}
   \rho(\mathbf{z},\lambda) = \frac{[E_{tot}-H({ \mathbf{z}},\lambda) ]^{C-1}_+}{\int d  \mathbf{z} [E_{tot}-H({ \mathbf{z}},\lambda) ]^{C-1}_+ }
 \label{eq:rho-general-case}
\end{equation}
if and only if the specific heat of the bath $C$ is energy-independent.
\end{theo}
Here $C$ is the microcanonical specific heat of the bath, i.e.:
\begin{equation}
C(E_B):= \left(\frac{\partial}{\partial E_B} T_B(E_B) \right)^{-1}
\end{equation}
where $E_B$ is the energy, and $ T_B(E_B) := \Phi_B(E_B)/\Omega_B (E_B)$ is the microcanonical temperature expressed in terms of the phase space volume $\Phi_B(E_B)$ of the bath, with energy below $E_B$, and the bath density of states $\Omega_B(E_B)={\partial \Phi_B(E_B)}/{\partial E_B}$.
In the following of this work we restrict ourselves to the case of energy independent, positive specific heat of the bath, abbreviated as $C(E_B):= C > 0$.

Note that the pdfs in Eq. (\ref{eq:rho-general-case}) are parametrized via the total system energy $E_{tot}$. It is however convenient to parametrize the pdfs via a property that pertains to the subsystem only, e.g., its average energy $U$. This is accomplished by writing $E_{tot}= U + C T$ (here $C T$ represents the average energy of the bath), substituting this expression in (\ref{eq:rho-general-case}), and imposing that $U = \int d \textbf{z} H( \textbf{z} ;\lambda)   \rho(\mathbf{z},\lambda)$. This leads to solving the following equation for $T$, given the average energy $U$ and $\lambda$
\begin{equation}
\frac{\int d\textbf{z} H( \textbf{z} ;\lambda) \left[ 1 - (H({ \textbf{z}};\lambda)-U)/({C T})\right]^{C-1}_+}
	{\int d\textbf{z} \left[ 1 - (H({ \textbf{z}};\lambda)-U)/(C T)\right]^{C-1}_+ }
	 =  U
	 \label{eq:<H>=U}
\end{equation}
We shall denote the value of $T$ that satisfies Eq. (\ref{eq:<H>=U}) for given $U$ and $\lambda$ as $T(U,\lambda)$  (in Appendix \ref{app:B} we prove that a solution $T(U,\lambda)$ always exists).
With this function at hand we can parametrize the pdfs in Eq. (\ref{eq:rho-general-case}) via the subsystem average energy $U$ and recast them in the form:
\begin{equation}
  \rho_{C}(\textbf{z};U,\lambda) = \frac{\left[ 1 -
(H({ \textbf{z}};\lambda)-U)/(C T(U,\lambda)) \right]^{C-1}_+}{N_{C}(U,\lambda) }
\label{eq:rho-c}
\end{equation}
where $N_{C}(U,\lambda)$ is the normalization:
\begin{equation}
N_{C}(U,\lambda)  = \int d\textbf{z} \left[ 1 -
(H(\textbf{z};\lambda)-U)/({C T(U,\lambda)})\right]^{C-1}_+
\label{eq:Nc}
\end{equation}

As discussed in Appendix \ref{app:B} it is not always possible to invert $T(U,\lambda)$.
For sake of simplicity, in this work we shall assume that $T(U,\lambda)$ is invertible with respect to the argument $U$. This means that we could also choose $T$ as an independent parameter and express $U$ as a function of $T$ and $\lambda$. Thus we are free to choose between two possible parameterizations: a microcanonical-like parameterization (or $U$-parameterization), and a canonical-like parameterization (or $T$-parameterization) \cite{note3}.

We shall refer to the numerator in Eq (\ref{eq:rho-c}) as to a ``generalized Boltzmann factor''. It is important to stress that a factor of the type $\left[ 1 -
{(H({ \textbf{z}};\lambda)-U)}/({C T})\right]^{C-1}_+$ is a generalized Boltzmann factor only if $T=T(U,\lambda)$, in agreement with Eq. (\ref{eq:<H>=U}).

\subsection*{Remark}
By expressing the specific heat $C$ as $C=1/(1-q)>0$ one recognizes that the pdf in Eq. (\ref{eq:rho-c}) is the Tsallis escort pdf of index $q$ with $q<1$  \cite{CampisiPHYSA2}.
Note that these \emph{do not} exhibit heavy tails but rather have a \emph{faster than exponential} decay with a finite cutoff occurring at the energy $U+CT=E_{tot}$. The physical meaning of this cutoff energy is that the system's energy cannot be larger than the total energy. 

\subsection*{Properties}
\subsubsection*{Equipartition}
The following equipartition theorem holds for the Finite Bath Statistics in (\ref{eq:rho-c}) \cite{CampisiPHYSA2}:
\begin{equation}
\langle \: p_i \frac{\partial H}{\partial p_i } \rangle = T(U,\lambda)
\label{eq:eq-part}
\end{equation}
where $\langle \cdot \rangle$ denotes average over $\rho_C$ in Eq. (\ref{eq:rho-c}), $p_i$ is one of the momenta and repeated indices are not summed. Eq. (\ref{eq:eq-part}) says that $T(U,\lambda)$ can be interpreted as the temperature of the system.

\subsubsection*{Heat Theorem}
The Finite Bath Statistics provides a \emph{mechanical model of thermodynamics} \cite{Gallavotti}, meaning that the temperature $T$, the external parameter $\lambda$, its conjugated generalized force $f_\lambda$, and the average energy $U$ are related in such a way as to satisfy the \emph{heat theorem} \cite{Campisi07}:
\begin{equation}
\frac{dU+f_\lambda d\lambda}{T} = \textit{ \emph{exact differential} }
\end{equation}
where $f_\lambda$ is defined in the usual way as:
\begin{equation}
f_\lambda =  - \langle \frac{\partial H}{\partial \lambda } \rangle
\end{equation}
This property is an important one because it allows to determine the thermodynamic entropy associated with the Finite Bath Statistics by finding the integral of the exact differential. This is given by \cite{Campisi07}:
\begin{equation}
S_{C}(U,\lambda) = \ln N_{C}(U,\lambda)\;.
\label{eq:logNC}
\end{equation}

\subsubsection*{Interpolation}
The pdfs in Eq. (\ref{eq:rho-c}) interpolate between canonical and microcanonical ensembles. Using the limits of infinite and null specific heat $C$, i.e.,
\begin{eqnarray}
&\lim_{C \rightarrow \infty}& \left[1+\frac{x}{C} \right]^{C-1}_+ = \lim_{C \rightarrow \infty} \left[1+\frac{x}{C} \right]^{C}_+ = e^x; \\
&\lim_{C \rightarrow 0}& \left[1+\frac{x}{C} \right]^{C}_+ \quad =  \theta(x); \\
&\lim_{C \rightarrow 0}&  \left[1+\frac{x}{C} \right]^{C-1}_+ = \delta(x)
\label{eq:limits}
\end{eqnarray}
one recovers the canonical and microcanonical pdfs \cite{note4}:
\begin{eqnarray}
\lim_{C \rightarrow \infty}  \rho_{C}(\textbf{z};T,\lambda) &=& \frac{e^{-H({\scriptsize \textbf{z}};\lambda)/T}}{Z(T,\lambda)} \\
\lim_{C \rightarrow 0}  \rho_{C}(\textbf{z};U,\lambda) &=& \frac{\delta(U-H(\textbf{z};\lambda))}{\Omega(U,\lambda)}, \label{eq:micLim}
\end{eqnarray}
respectively \cite{Campisi06.2}. The microcanonical normalization ${\Omega(U,\lambda)}$ is the system density of states.
Likewise one has, for the normalization, the following limits \cite{Campisi06.2}:
\begin{eqnarray}
\lim_{C \rightarrow \infty} N_{C}(T,\lambda) &=&  \int d \textbf{z} e^{-(H({\scriptsize \textbf{z}};\lambda)-U)/T} \nonumber \\
&=& e^{U/T}Z(T,\lambda) \label{eq:Z} \\
\lim_{C \rightarrow 0}  N_{C}(U,\lambda) &=& \int_{H({\scriptsize \textbf{z}};\lambda) \leq U} d \textbf{z} = \Phi(U,\lambda) \label{eq:Phi}
\end{eqnarray}
The quantity $\Phi(U,\lambda)$ is the volume of system phase space with energy below $U$. The density of states is related to $\Phi$ via a partial derivative $\Omega = {\partial \Phi}/{\partial U}$.
By taking the logarithm one recovers canonical and microcanonical entropies; i.e.,
\begin{eqnarray}
\lim_{C \rightarrow \infty} S_{C}(T,\lambda) &=&  {U \over T} +\ln Z(T,\lambda) \label{eq:lnZ} \\
\lim_{C \rightarrow 0}  S_{C}(U,\lambda) &=&\ln \Phi(U,\lambda) \label{eq:lnPhi}
\end{eqnarray}

\section{\label{sec:FT}The fluctuation theorem}
Consider an ensemble of systems distributed according to Eq (\ref{eq:rho-c}). Assume the system being decoupled from its bath and that it is acted upon by an external force that changes the external parameter $\lambda$ according to some prescribed protocol $\lambda(t)$ executed between times $t_0$ and $t_f$. The probability density that the external force does a certain work $W$ on the system in that interval of time reads:
\begin{equation}
\begin{split}
p_{t_f,t_0}^{C,U}(W) &:= N_{C,0}^{-1}(U) \\
& \times \int d  \textbf{z}_0  \delta(H_f( \textbf{z}_f)-H_0(\textbf{z}_0)-W) \\
& \times \left[  1- \frac{H_0(\textbf{z}_0)-U}{C T_0(U)}\right]^{C-1}_+
\label{eq:defPDFwork}
\end{split}
\end{equation}
where $\textbf{z}_f=\textbf{z}(t_f,t_0,\textbf{z}_0)$ is the solution of Hamilton's equation with initial condition $\textbf{z}_{0}$. For simplicity of notation we drop the variable $\lambda$ in all quantities that depend on it, and replace it with a subscript $0$ or $f$, depending on whether the quantity is taken at values of $\lambda$ equal to $\lambda(t_0)$ or $\lambda(t_f)$, e.g., $H_0(\textbf{z})=H(\textbf{z},\lambda(t_0))$, $T_0(U)=T(U,\lambda(t_0))$.
By making the change of variables from $\textbf{z}_0 \rightarrow \textbf{z}_f$ with a unitary Jacobian, one obtains
\begin{equation}
\begin{split}
N_{C,0}(U) p_{t_f,t_0}^{C,U}(W)=
\int d   \textbf{z}_f \delta(H_0  ( \textbf{z}_0)-H_f(\textbf{z}_f)+W) \\ \times
 \left[ 1- \frac{H_f(\textbf{z}_f)-(U+W)}{C T_0(U)}\right]^{C-1}_+
\end{split}
\label{eq:npBF}
\end{equation}
where now $\textbf{z}_0=\textbf{z}(t_0,t_f,\textbf{z}_f)$, is the solution of Hamilton's equation with $\textbf{z}_f$ as initial condition and time running backward.
Note that the second term in the integrand \emph{is not} a generalized Boltzmann factor because in general it does not satisfy Eq. (\ref{eq:<H>=U}).
However for any $\delta T$ one can rewrite the previous equation as:
\begin{equation}
\begin{split}
N_{C,0}(U) p_{t_f,t_0}^{C,U}(W) & = \left( \frac{T_0(U)+ \delta T}{T_0(U)}\right)^{C-1} \\
& \times \int d \textbf{z}_f \delta(H_0( \textbf{z}_0)-H_f(\textbf{z}_f)+W)   \\
 &\times \left[ 1- \frac{H_f(\textbf{z}_f)-(U+W-C\delta T)}{C(T_0(U)+\delta T)}\right]^{C-1}_+
\end{split}
\end{equation}
We now choose $\delta T$ as the solution of the following integral equation:
\begin{equation}
\frac{\int d\textbf{z} H_f(\textbf{z}) B(\textbf{z},U,W,\delta T) }{\int d\textbf{z} B(\textbf{z},U,W,\delta T)}   = U+W-C\delta T
\label{eq:deltaT}
\end{equation}
where, for convenience, we introduce the notation
\begin{equation}
B(\textbf{z},U,W,\delta T) := \left[ 1- \frac{H_f({\textbf{z}})-(U+W-C\delta T)}{C(T_0(U)+\delta T)}\right]^{C-1}_+;
\end{equation}
or, equivalently as a solution of:
\begin{equation}
T_0(U)+\delta T = T_f(U+W-C\delta T)
\label{eq:deltaT2}
\end{equation}
Then, we find:
\begin{equation}
\begin{split}
N_{C,0}(U) p_{t_f,t_0}^{C,U}(W) &= \left(\frac{T_f (U+W-C\delta T)}{T_0(U)}\right)^{C-1} \\
& \times
\int d \textbf{z}_f \delta(H_0( \textbf{z}_0)-H_f(\textbf{z}_f)+W) \\
& \times
 \left[ 1- \frac{H_f(\textbf{z}_f)-(U+W-C\delta T)}{T_f (U+W-C\delta T)}\right]^{C-1}_+
 \end{split}
\end{equation}
where the second term of the integrand is the Boltzmann factor of the pdf $\rho_{C}(\textbf{z};U+W-C\delta T,\lambda(t_f))$. The integral is the product of $N_{C,f}(U+W-C\delta T)$ and the probability ${p_{t_0,t_f}^{C,U+W-C\delta T}(-W)}$ that the force performs the work $-W$ when the protocol is run backward and the system is initially in the state $\rho_{C}(\textbf{z};U+W-C\delta T,\lambda(t_f))$.

Therefore the following fluctuation theorem is obtained:
\begin{equation}
\frac{p_{t_f,t_0}^{C,U}(W)}{p_{t_0,t_f}^{C,U_f}(-W)} = \left(\frac{T_f}{T_0}\right)^{C-1} \frac{N_{C,f}(U_f)}{N_{C,0}(U)},
\label{eq:genFT}
\end{equation}
where,
\begin{eqnarray}
U_f &:=& U+W-C \delta T \label{eq:Uf}\\
T_f &:=& T_f(U_f)
\end{eqnarray}
Using Eq. (\ref{eq:logNC}), Eq. (\ref{eq:genFT}) can be rewritten in terms of entropy as:
\begin{equation}
\frac{p_{t_f,t_0}^{C,U}(W)}{p_{t_0,t_f}^{C,U_f}(-W)} = \left(\frac{T_f}{T_0}\right)^{C-1} \exp [\Delta S_{C}^{f,0}(U,W)]
\label{eq:genFT2}
\end{equation}
where $\Delta S_{C}^{f,0}(U,W)=S_{C,f}(U_f)-S_{C,0}(U)$.

The Finite Bath Fluctuation Theorem of Eq.  (\ref{eq:genFT2}) allows to calculate the ratios
of probability of work done on the system when it is driven arbitrarily away from
equilibrium during the action of the forward and backward protocol, in terms of
equilibrium properties such as entropy and temperature.

\subsection*{Recovering known special cases}
\subsubsection*{Limit of microcanonical ensemble}
In the limit $C \rightarrow 0$ Eq. (\ref{eq:npBF}) becomes (using the formula $\delta(ax)=a^{-1}\delta(x)$, and Eqs. (\ref{eq:limits}) and (\ref{eq:Phi}))
\begin{equation}
\begin{split}
&\Phi_0(U) p_{t_f,t_0}^{C,U}(W)=
T_0(U) \\
&\times \int d \textbf{z}_f \delta(H_0( \textbf{z}_0)-H_f(\textbf{z}_f)+W) \delta (H_f(\textbf{z}_f)-(U+W) )
\end{split}
\end{equation}
Using the microcanonical equipartition theorem \cite{Khinchin}
$T(U,\lambda)={\Phi(U,\lambda)}/{\Omega(U,\lambda)}$, one recovers
the microcanonical fluctuation theorem \cite{Cleuren:2006le,Talkner:2008gb}:
\begin{equation}
\frac{p_{t_f,t_0}^{0,U}(W)}{p_{t_0,t_f}^{0,U+W}(-W)} =  \frac{\Omega_{f}(U+W)}{\Omega_{0}(U)}.
\end{equation}
Alternatively one can take the limit $C \rightarrow 0$ of Eq. (\ref{eq:genFT})
directly and obtain the expression ${T_0(U) \Phi_f(U+W)}/({T_f(U+W) \Phi_0(U)})$,
which reduces to the previous one by virtue of the microcanonical equipartition theorem.

\subsubsection*{Limit of canonical ensemble}
Likewise, using the $T$ parameterization, it can be seen that, in the limit $C \rightarrow \infty$ Eq. (\ref{eq:npBF}) becomes
\begin{equation}
\begin{split}
Z_0(T) p_{t_f,t_0}^{C,T} & (W) = e^{W/T} \\
& \times
\int d \textbf{z}_f \delta(H_0( \textbf{z}_0)-H_f(\textbf{z}_f)+W)e^{- {H_f({\scriptsize \textbf{z}_f)}}/{T} }
\label{eq:canFT}
\end{split}
\end{equation}
One thus obtains the fluctuation theorem for the canonical ensemble of Crooks  \cite{Crooks:1999fq,talkner2007}:
\begin{equation}
\frac{p_{t_f,t_0}^{\infty,T}(W)}{ p_{t_0,t_f}^{\infty,T}(-W) } =  \frac{Z_{f}(T)}{Z_{0}(T)} e^{W/T}
\label{eq:canFT2}
\end{equation}

\section{\label{sec:example}Example: A 2D gas of hard disks}
In this section we illustrate the Finite Bath Fluctuation Theorem  by applying it to a system composed of $n+1$ elastically colliding hard disks in a 2-dimensional box with perfectly reflecting walls. One disk will be our system of interest, whereas the remaining $n$ ones will form the bath. We assume that the disks do not have rotational degrees of freedom. As shown in the Appendix \ref{app:A}, the specific heat is given in this case by $C={dn}/{2}$ where $d$ is the number of translational degrees of freedom of each disk. In this case $d=2$, hence $C=n$. Note the fact that $C$ does not depend on energy.
\subsection*{The probability density function}
\begin{figure}
\begin{center}
\includegraphics[width=8cm]{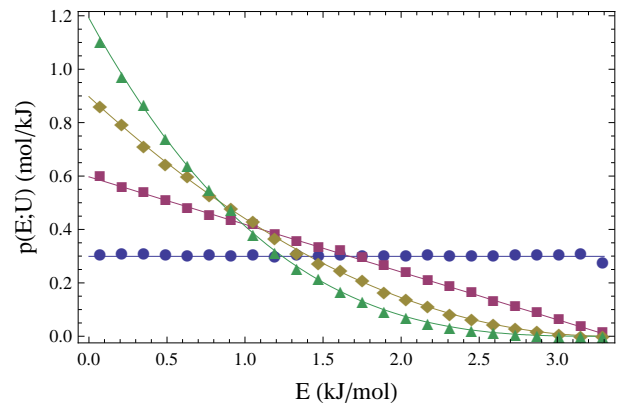}
 \caption{(Color online) Energy probability density function (pdf) for a 2D hard disk of radius $r = 1$ nm and mass $M=2$ amu, in a bath composed of 1 ($\bullet$), 2 ($\blacksquare$), 3 ($\blacklozenge$), 4($\blacktriangle $) other identical disks. The dots represent histograms of properly normalized relative frequencies from numerical simulations. All simulations were carried out for the same total energy $E_{tot}=3.3469$ kJ/mol, which corresponds to measured average energies of the disk of interest $U_1=1.67166$ kJ/mol, $U_2=1.11644$  kJ/mol, $U_3=0.835784$ kJ/mol, $U_4=0.671521$ kJ/mol. The solid lines represent the pdf predicted by the theory (Eq. (\ref{eq:p(E)})) for the measured average energies $U_i, i=1..4$.}
\label{fig:PDF}
\end{center}
\end{figure}

The energy of the system of interest is simply its kinetic energy; i.e.,
\begin{equation}
H(p_x,p_y;M)= \frac{p_x^2+p_y^2}{2M},
\label{eq:2DHam}
\end{equation}
which fluctuates permanently due to the collisions with the bath's particles. According to Eq. (\ref{eq:rho-c}), the probability that the disk has a given momentum $(p_x,p_y)$ is given by
\begin{equation}
\begin{split}
  \rho_{C}&(p_x,p_y;U,M) \\
  &= {N_{C}^{-1}(U,M)}
  {\left[ 1 -
\frac{(p_x^2+p_y^2)/(2M)-U}{CT(U,M)}\right]^{C-1}_+}
\label{eq:2D-rho}
\end{split}
\end{equation}
We consider the mass of the disk $M$ as an external parameter that can be changed at will in the course of time according to pre-specified protocols. The function $T(U,M)$ has to be computed via Eq. (\ref{eq:<H>=U}). In general, the solution of Eq. (\ref{eq:<H>=U}) with a purely kinetic Hamiltonian with $s$ translational degrees of freedom gives the usual equipartition of energy \cite{CampisiPHYSA2}:
$
T(U,M) = {2U}/{s}
$.
In the specific case of Eq. (\ref{eq:2DHam}) $s=2$, hence
\begin{equation}
T(U,M)=U \;,
\label{eq:T(U)}
\end{equation}
and
\begin{equation}
\begin{split}
  \rho_{C}&(p_x,p_y;U,M) \\
  & =  {N_{C}^{-1}(U,M)}
{\left[ 1 -
\left(\frac{p_x^2+p_y^2}{2M}-U\right) /({CU}) \right]^{C-1}_+}
\label{eq:2D-rho2}
\end{split}
\end{equation}
Using Eq. (\ref{eq:Nc}), with (\ref{eq:2D-rho2}) gives
\begin{equation}
N_{C}(U,M)= 2\pi A [1+C^{-1}]^{C}M U
\label{eq:Nc2D}
\end{equation}
where $A$ is the reduced volume (i.e., area in this 2-dimensional case) of the box (see the Appendix \ref{app:A} for the definition of reduced volume).
From Eq. (\ref{eq:2D-rho2}), one obtains the pdf of energy $E$ of the disk:
\begin{equation}
p(E;U) = {U^{-1}[1+C^{-1}]^{-C}}{\left[ 1 -
\frac{(E-U)}{C U}\right]^{C-1}_+}.
\label{eq:p(E)}
\end{equation}
Interestingly, the energy pdf does not depend on the mass $M$. In Fig. \ref{fig:PDF} we compare Eq. (\ref{eq:p(E)}), with the result of various numerical simulations with $C=1,2,3,4$. Note that for $C=1$ the distribution is flat, for $C=2$ it is linear, for $C=3$ it is quadratic etc...
In view of theorem 1, the impressive agreement between theory and numerics corroborates the validity of the assumed ergodic hypothesis for this model system.
Similar simulations have been reported in \cite{Adib04} for a 1-dimensional harmonic oscillator coupled to a bath of $n$ 1-dimensional quartic oscillators. In that case the density of states of the bath is proportional to $E^{(3n-2)/4}$, and accordingly the specific heat, $C=(3n+2)/4$, is energy independent.
\subsection*{Analytical test of the Finite Bath Fluctuation Theorem }
Consider a protocol $M(t)$ that changes the mass of the disk from the value $M_0=M(t_0)$ to $M_f=M(t_f)$. According to the general assumption of our derivation, the system is decoupled from the bath during the action of the protocol. We are interested in checking the validity of Eq. (\ref{eq:genFT}). To this end we need to compute the forward pdf of work, $p_{t_f,t_0}^{C,U}(W)$, the backward pdf of work $p_{t_0,t_f}^{C,U_f}(-W)$, and the starting average energy of the backward protocol $U_f$, given the starting average energy of the forward protocol $U$. Solving Eq. (\ref{eq:deltaT2}) with Eq. (\ref{eq:T(U)}) (note that Eq. (\ref{eq:T(U)}) does not depend on the value of $M$, hence $T_f(U)=T_0(U)=U$) we arrive at:
\begin{equation}
\delta T = {W}/({1 +C})
\label{eq:deltaT2D}
\end{equation}
hence from Eq. (\ref{eq:Uf}) we obtain
\begin{equation}
U_f = T_f= U+{W}/({1+C}).
\label{eq:Uf2D}
\end{equation}
Using Eq. (\ref{eq:Nc2D}) with (\ref{eq:Uf2D}) we obtain the normalizations of the equilibrium pdfs with average energy and external parameters $(U,M_0)$ and $(U_f,M_f)$ respectively:
\begin{eqnarray}
N_{C,0}(U)&=& 2\pi A [1+C^{-1}]^{C}M_0 U \label{eq:norm2D0} \\
N_{C,f}(U_f)&=& 2\pi A [1+C^{-1}]^{C}M_f \left(U+ \frac{W}{1+C}\right) \label{eq:norm2Df}
\end{eqnarray}
Using (\ref{eq:Uf2D},\ref{eq:norm2D0},\ref{eq:norm2Df}) we find:
\begin{equation}
\left(\frac{T_f}{T_0}\right)^{C-1} \frac{N_{C,f}(U_f)}{N_{C,0}(U)} = \left(\frac{U_f}{U}\right)^{C}\frac{M_f}{M_0}
\label{eq:rhsFTspec}
\end{equation}
From Eq. (\ref{eq:defPDFwork}) we have:
\begin{equation}
\begin{split}
p_{t_f,t_0}^{C,U}(W) & := N_{C,0}^{-1}(U)A  \\
& \times
\int dp_x dp_y  \delta \left(\frac{p_x^2+p_y^2}{2M_f}-\frac{p_x^2+p_y^2}{2M_0}-W\right)\\
& \times
\left[ 1 -
\left(\frac{p_x^2+p_y^2}{2M_0}-U\right) /({CU})\right]^{C-1}_+
\end{split}
\end{equation}
where we use the fact that the momentum $(p_x,p_y)$ is a constant of motion.
By applying the change of variable $E=(p_x^2+p_y^2)/(2M_0)$, and employing Eq. (\ref{eq:norm2D0}) we obtain:
\begin{equation}
\begin{split}
p_{t_f,t_0}^{C,U}(W)  &= U^{-1}[1+C^{-1}]^{-C}\frac{M_f}{|M_0-M_f|} \\
& \times\left[ 1 -
\left( \frac{M_f}{M_0-M_f}W -U\right)/({C U})\right]^{C-1}_+
\label{eq:p+}
\end{split}
\end{equation}
Similarly one finds the backward pdf of work
\begin{equation}
\begin{split}
p_{t_0,t_f}^{C,U_f}(-W) &= U_f^{-1}[1+C^{-1}]^{-C}\frac{M_0}{|M_f-M_0|}\\
& \times \left[ 1 -
\left( \frac{M_0}{M_0-M_f}W -U_f\right)/({C U_f})\right]^{C-1}_+
\label{eq:p-}
\end{split}
\end{equation}
Taking the ratio of Eq. (\ref{eq:p+}) and Eq. (\ref{eq:p-}) we obtain:
\begin{equation}
\frac{p_{t_f,t_0}^{C,U}(W)}{p_{t_0,t_f}^{C,U_f}(-W)} = \left(\frac{U_f}{U}\right)^{C}\frac{M_f}{M_0}
\label{eq:lhsFTspec}
\end{equation}
By comparison with Eq. (\ref{eq:rhsFTspec}) we see that the Finite Bath Fluctuation Theorem of Eq. (\ref{eq:genFT}) is satisfied.

\subsection*{Numerical check of the Finite Bath Fluctuation Theorem }
\begin{figure}
\begin{center}
\includegraphics[width=8cm]{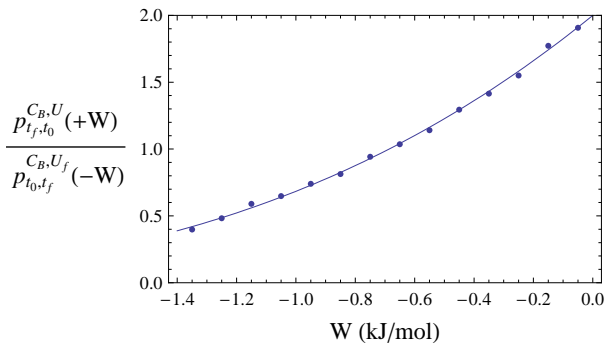}
 \caption{(Color online) Comparison between the numerical values (dots) and the theoretical expression in Eq. (\ref{eq:lhsFTspec})  (continuous line) of ${p_{t_f,t_0}^{C,U}(W)}/{p_{t_0,t_f}^{C,U_f}(-W)}$ for a 2D hard disk of mass $M_0=2$ amu in a bath composed of 3 hard disks of the same mass. The initial energy is $U=0.831447$ kJ/mol and the protocol doubles the mass of the disk.}
\label{fig:numcheck}
\end{center}
\end{figure}
In order to check numerically the validity of Eq. (\ref{eq:lhsFTspec}) we simulated the forward work pdf $p_{t_f,t_0}^{C,U}(W)$ for a bath of $n=C$ 2D disks, a given value of $U$ and a protocol that changes the mass of the disk from $M_0$ to $M_f=2M_0$. The pdf for the numerical work  is calculated as follows. We first run a simulation of the motion of the disk with fixed $U$ and $M_0$. We then construct a histogram that counts the number of occurrences of energy in the intervals $I_n=[E_n-\Delta E/2, E_n+\Delta E/2$) for a certain $\Delta E$ (in our simulations, typically, $\Delta E=0.$1 kJ/mol, for a total of about 20 intervals, and the histogram counts a total of about $10^5$ events). This provides us with the starting statistics.
At this point, we note that, independent of the functional form of $M(t)$, acting the protocol on a particle with energy $E$ gives with probability 1 the work $W=E (M_0-M_f)/{M_f}$. The reason is that the time dependent system Hamiltonian $(p_x^2+p_y^2)/(2M(t))$ generates the following equation of motion for the momenta: $\dot{p_x}=\dot{p_y}=0$. Hence $E(t_f)=(p_x^2+p_y^2)/(2M(t_f))= E(t_0)M_0/M_f$ regardless of the details of the protocol.
So we immediately obtain a count of work belonging to the intervals $J_n=[W_n-\Delta W, W_n+\Delta W$), where $W_n=E_n(M_0-M_f)/{M_f}$ and $\Delta W=\Delta E(M_0-M_f)/{M_f}$. After proper normalization, this yields a histogram, labeled
\begin{equation}
h_{t_f,t_0}^{C,U}(n) \nonumber
\end{equation}
that provides a numerical estimate for  $p_{t_f,t_0}^{C,U}(W)$. Next, for each $n$, we simulate the motion of the disk with fixed parameters, $M_f=2M_0$ and $U_n = U+{W_n}/({C+1})$, and compute $n$ different histograms for the backward probabilities $h_{t_0,t_f}^{C,U_n}(k)$ in the same way as the forward histogram was computed. By selecting the $k=n$ value from each of the backward histograms and collecting them to form the new histogram
\begin{equation}
h_{t_0,t_f}^{C,U_n}(n) \nonumber
\end{equation}
we obtain a numerical estimate for  $p_{t_0,t_f}^{C,U_f}(-W)$. Finally, we compute the ratios ${h_{t_f,t_0}^{C,U}(n)}/{h_{t_0,t_f}^{C,U_n}(n)}$.

These ratios are depicted in Figure \ref{fig:numcheck} along with the theoretical values given by Eq. (\ref{eq:lhsFTspec}). The figure shows excellent agreement between analytical theory and numerical experiment. The visible differences are within the statistical errors.
Note that, for the forward protocol, where the mass is increased by a factor 2, the work can only be negative, and vice-versa for the backward protocol. Therefore, the graph shows only the negative values of nonequilibrium work $W$.

\section{\label{sec:discussion}Discussion}

\subsection*{Physical meaning of $\delta T$}
The basic quantity that enters the Finite Bath Fluctuation Theorem, and marks a distinction with the canonical fluctuation theorem of Crooks (\ref{eq:canFT2}), is the quantity $\delta T$, defined formally as the solution of Eq. (\ref{eq:deltaT2}).
This quantity enters in the definition of $U_f$ and $T_f$. What is the physical meaning of these quantities?
The hard sphere gas example turns useful in addressing this question. Calculations analogous to those leading to Eq. (\ref{eq:deltaT2D}) show that for a gas of hard spheres with a total of $s$ degrees of freedom, in contact with a bath with a specific heat $C$, it is:
\begin{equation}
\delta T = W/C_{tot}
\end{equation}
where $C_{tot}$ is the total specific heat of the system+bath compound system: $C_{tot}:=s/2+C$.
This $\delta T$ is therefore the increment of temperature that would result if, after having injected the energy $W$ in the system of interest this is brought back into contact with the bath, and the compound system is let reach thermal equilibrium. Recall that  during the forcing protocol we assumed that system and bath are decoupled. We shall refer to this process as to the \emph{re-thermalization}. After system and bath have re-thermalized, the extra energy $W$, initially stored in the system, will be shared between system and bath according to the ratio of the respective specific heats. In particular the bath gets the energy $Q=C\delta T$, which is indeed the heat that flows from the system to the bath during re-thermalization. Accordingly the system looses this amount of energy, and its change in energy becomes $\Delta U = W-Q$, in agreement with  the first law of thermodynamics. This means that $U_f$ represents the average energy of the system after the re-thermalization. To summarize: (a) The system is first in thermal contact with the bath. Its average energy is $U_i$ and the temperature is $T_i$. (b) the system is decoupled from the bath and the forcing protocol is acted on it. As a result, the energy $W$ is injected in the system with a certain probability density $p^{C,U}_{t_f,t_0}(W)$. (c) The system (carrying the extra energy $W$), and bath (still at temperature $T_i$) are now allowed to re-thermalize. During re-thermalization the heat $C\delta T$ flows in the bath, the system reaches the average energy $U_f$, and the new temperature $T_f$ is reached in the compound system.

Remarkably, the temperature change $\delta T$ vanishes in the canonical case: $\lim_{C\rightarrow \infty} \delta T=0$. However it is $\lim_{C\rightarrow \infty} C \delta T=W$, meaning that the whole extra energy $W$ injected in the system, flows into the bath during re-thermalization. However this does not affect its temperature (i.e., $T_i=T_f$), the specific heat being infinite in the canonical case. Therefore the term $T_f/T_0$ does not appear in the canonical fluctuation theorem of Crooks. In fact the latter gives information about the free energy difference of two states with different parameter values, but \emph{same} temperature. This is a much more fortunate situation as compared to the finite bath and microcanonical fluctuation theorems, in the sense that, in the canonical case, one should not bother to start the backward process from the ``target" temperature $T_f$ (which depends on $W$), but simply starts it from the same temperature as that of the forward process.

\subsection*{Implications for the second law of thermodynamics}
From the canonical fluctuation theorem of Crooks, one obtains, after proper algebraic manipulations, and integration over $W$, the integral form of the fluctuation theorem, namely the Jarzynski equality $\langle e^{-\beta W}\rangle= e^{-\beta \Delta F}$ \cite{Jarzynski1997}, which implies the second law in the form $\langle W\rangle \geq \Delta F$.
A similar integral equation can be obtained for the Finite Bath Fluctuation Theorem too. It reads:
\begin{equation}
\mathcal{N} \langle T_f^{C-1}e^{S_f(U_f)}\rangle= T_0^{C-1}e^{S_0(U_0)}
\end{equation}
where, 
\begin{equation}
\mathcal{N}:= \int p_{t_0,t_f}^{U_f}(W)dW
\label{eq:intFlucTheo}
\end{equation}
and $\langle \cdot \rangle$ denotes average over the normalized distribution $q_{t_0,t_f}(W):=p_{t_0,t_f}^{U_f}(W)/\mathcal{N}$. Eq. (\ref{eq:intFlucTheo}) generalizes both the canonical Jarzynski equality and the microcanonical \emph{Entropy-from-work theorem}  \cite{Talkner:2008gb,CampisiPRE08a}.
Note that, as for the Entropy-from-work theorem, in general it is $\mathcal{N} \neq 1$ because the energy $U_f$ in Eq. (\ref{eq:intFlucTheo}) is a function of $W$ (see Eq. \ref{eq:Uf}). As pointed out already in \cite{CampisiPRE08a}, this prevents obtaining the second law directly from the integral form of the fluctuation theorem.

Nevertheless the validity of the second law of thermodynamics for a driving protocol acting on a system that is initially thermalized with a finite bath,  can be proved directly without invoking the Finite Bath Fluctuation Theorem. To this end it is sufficient to recall the content of two theorems which have been recently reported in the literature \cite{CampisiPRE08b,CampisiSHPMP08,AllahverdyanPRE05}. According to these theorems, the second law of thermodynamics, in either the minimal work principle form, or the entropy increase form of Clausius, is obeyed whenever the initial phase space pdf $\rho(\mathbf{z})$ is a \emph{decreasing} function of energy, namely $\rho(\mathbf{z})\geq \rho(\mathbf{z}')$, for every $\mathbf{z},\mathbf{z}'$ such that $H(\mathbf{z})\leq H(\mathbf{z}')$. This condition is obeyed by the Finite Bath Statistics, if the condition $C\geq 1$ is met (see Eq. \ref{eq:rho-general-case}). In this regard we notice that this condition only is violated in the extremal case when the bath consists of a single degree of freedom (in which case it is $C=1/2$), or if there is no bath at all ($C=0$, microcanonical case). 

The Crooks fluctuation theorem (\ref{eq:canFT2}) can be seen as a statement according to which the probability of doing a certain negative work $-W$ during the backward protocol is \emph{exponentially suppressed} with respect to the probability of doing the positive work $W$, in the forward protocol. For a cyclic protocol, this says that it is exponentially more probable to spend energy, rather harvesting it, in agreement with the Kelvin postulate (i.e., no energy extraction from a cyclic process). A similar situation occurs for the Finite Bath Fluctuation Theorem, with the exponential suppression being replaced by a power-law suppression. To exemplify this, consider again the gas of $N$ hard spheres in $d$ dimensions. Imagine the protocol consists of changing the volume of the box that contains the gas from $V_0$ to $V_f$. Straightforward calculations lead the following form of the Finite Bath Fluctuation Theorem
\begin{equation}
\frac{p_{t_f,t_0}^{C,U}(W)}{p_{t_0,t_f}^{C,U_f}(-W)}=\left(\frac{V_f}{V_i}\right)^{N/d} \left(1+\frac{W}{C_{tot}T_0}\right)^{C_{tot}-1}
\label{eq:FlucTheo3}
\end{equation}
 where it is evident that the power-law term $(1+W/(C_{tot} T_0))^{C_{tot}-1}$ becomes the exponential term appearing  in the Crooks theorem (\ref{eq:canFT2}) for very large $C$ ($C_{tot}=C+dN/2$ becomes very large for very large $C$).

\section{\label{sec:conl}Conclusions}

We devised a Finite Bath Fluctuation Theorem that gives information about the probability of work on systems that have been thermalized with a finite heat bath. This corresponds to physical situations which are situated between the two ideal cases of absent bath (microcanonical ensemble) and infinite bath (canonical ensemble). The Finite Bath Fluctuation Theorem interpolates between microcanonical and canonical fluctuation theorems. It thus generalizes these theorems and reveals a common underlying mathematical structure.

The validity of the Finite Bath Statistics is illustrated by means of numerical simulations of a 2D gas of hard disks in a box with perfectly reflecting walls, see Figure \ref{fig:PDF}, and the validity of the Finite Bath Fluctuation Theorem is confirmed both analytically and numerically, cf. Figure \ref{fig:numcheck}, for our system.

Similarity and differences between the Finite Bath Fluctuation Theorem and the canonical and microcanonical fluctuation theorems have been discussed, as well as its interrelation with the second law of thermodynamics. In contrast with the canonical fluctuation theorem, two temperatures, instead of one, appear in the Finite Bath Fluctuation Theorem. The physical meaning of these two temperatures has been clarified by considering a re-thermalization process.

As shown in Sec. \ref{sec:FHBS}, the Finite Bath Statistics in  (\ref{eq:rho-c}) is a special instance of the general statistical formula according to which the bath density of states determines the shape of the system pdf.
Based on quasi-adiabatic perturbation theory of chaotic systems, Jarzynski \cite{JARZYNSKI:1995qd} found that a slow particle coupled to a small bath with fast chaotic degrees of freedom thermalizes and reaches a stationary pdf whose shape is dictated by the density of states of the bath. Our simulations provide an example that such behavior of the system pdf occurs even if there is no time-scale separation between system and bath. In any case, thermalization of the subsystem towards a pdf of the form in Eq. (\ref{eq:rho-c}) is expected only if the total system is ergodic.

An important assumption underlying our main finding is that we used a specific heat that is energy-independent: Whether a  Finite Bath Fluctuation Theorem exists also in the case of more realistic energy dependent specific heats remains an open challenge.

\section*{Acknowledgements}
Financial support by the German Excellence Initiative via the {\it
Nanosystems Initiative Munich} (NIM) and the Volkswagen
Foundation (project I/80424) is gratefully acknowledged.

\begin{thebibliography}{10}
\bibitem{note1}
Sometimes this pdf is expressed in the system energy space, $E$, rather than in its phase space $\mathbf{z}$. For example in Ref. \cite{Prosper93}, the energy pdf of the system is given as:
$P(E)= const (E_{tot}-E)^{3n/2-1} E^{3m/2-1}$. The term $E^{3m/2-1}$ comes from the system's density of states, where the system is assumed to be itself an ideal gas of $m$ particles. Eq. (\ref{eq:rho-special-case}) is more general in that the system is not assumed to be ideal.
\bibitem{note2}
In this work we adopt the convention of measuring temperature in units of energy. Thus $k_B$, the Boltzmann constant is equal to 1, and the specific heat is dimensionless.
\bibitem{note4}
In taking the canonical limit we use the $T$-parameterization.
\end{thebibliography}

\appendix
\section{\label{app:A}Specific heat of a bath of $n$ hard spheres}
Although straightforward, the calculation of the microcanonical specific heat of a gas of hard spheres is not discussed in statistical mechanics textbooks. We present this calculation below.

The Hamiltonian of a gas of $n$ $d$-dimensional hard spheres of radius $a$ reads:
\begin{equation}
H_B(\{ \overrightarrow{p_i}\} ,\{\overrightarrow{q_i}\})= \sum_{i=1}^{n}\frac{\overrightarrow{p_i}^2}{2m}+\sum_{i<j}V(|\overrightarrow{q_i} - \overrightarrow{q_j}|),
\label{eq:HardSphereH}
\end{equation}
where $ \overrightarrow{p_i} ,\overrightarrow{q_i}$ are the $d$-dimensional momentum and position vectors of the $i^{th}$ sphere, and 
\begin{equation}
 V(x)=
 \left\{
  \begin{array}{ll}
0 &  x \geq a \\
+\infty  &x<a
 \end{array}
 \right.
 \label{eq:V(x)}
\end{equation}
is the hard core interaction potential.
The phase space volume $\Phi_B$ with energy below $E_B$ becomes:
\begin{equation}
\begin{split}
\Phi_B(E_B)= & \int \prod_{i=1}^n d \overrightarrow{q_i} \int \prod_{i=1}^n d \overrightarrow{p_i} \\
& \times \theta\left(E_B-\sum_{i=1}^{n}\frac{\overrightarrow{p_i}^2}{2m}-\sum_{i<j}V(|\overrightarrow{q_i} - \overrightarrow{q_j}|\right),
\end{split}
\end{equation}
where each integral in $d \overrightarrow{q_i} $ is restricted to the region $\mathcal{V}$, of volume V, of the box. For values of $|\overrightarrow{q_i} - \overrightarrow{q_j}|$ smaller than $a$, the integrand vanishes, thus reducing the spatial integration domain to the region $\mathcal{M} \subset \mathcal{V}^n$ where \break $|\overrightarrow{q_i} - \overrightarrow{q_j}|> a $, for each couple $i,j$. In this region the interaction term is zero and one obtains:
\begin{equation}
\begin{split}
\Phi_B(E_B)= & V'^n \int \prod_{i=1}^n d \overrightarrow{p_i} \theta\left(E_B-\sum_{i=1}^{n}\frac{\overrightarrow{p_i}^2}{2m} \right),\\
\quad
\end{split}
\end{equation}
where $V'^n= \int_{\mathcal{M} } \prod_{i=1}^n d \overrightarrow{q_i}$, is independent of $E_B$. We shall refer to $V'$ as to the reduced volume. The integration over the momenta then yields \cite{Huang}:
\begin{equation}
\Phi_B(E_B)= A_{dn} (2m)^{dn/2} V'^n E_B^{dn/2}
\end{equation}
where $A_{N}:=\pi^{N/2}/\Gamma(N/2+1)$.
By differentiating $\Phi_B(E_B)$ with respect to $E_B$, one finally obtains the density of states of the gas of hard spheres:
\begin{equation}
\Omega_B(E_B)= A_{dn} (dn/2) (2m)^{dn/2} V'^N E_B^{dn/2-1}
\end{equation}
The only difference with the density of states of an ideal gas is that the actual volume $V$ is replaced by the reduced volume $V'$.
The temperature $T_B(E_B)=\Phi_B(E_B)/\Omega_B(E_B)$, is given by same formula as for the ideal gas, i.e.,  $T_B(E_B)=2E_B/(dn)$, and so is the specific heat, i.e., $C(E_B)=dn/2$.
For simplicity, in Eq. (\ref{eq:HardSphereH}) we neglected the spheres rotational degrees of freedom. These however would add to the total specific heat an energy independent contribution.

\section{\label{app:B}Existence and (non)uniqueness of solutions of Eq. (\ref{eq:<H>=U}) }
We prove that, given $U$ and $\lambda$, it is always possible to find a $T$ such that Eq. (\ref{eq:<H>=U}) is satisfied.
For this purpouse we define the function:
\begin{equation}
I_\lambda(U,T):=\int_0^{CT+U} de \Omega_\lambda(e) (e-U)(CT-e+U)^{C-1}
\end{equation}
which is continuous with respect to both $U$ and $T$.  The symbol $ \Omega_\lambda(e)$ denotes
the density of states of the Hamiltonian $H(\mathbf{z},\lambda)$.
Eq. (\ref{eq:<H>=U}) can be equivalently expressed as:
\begin{equation}
I_\lambda(U,T)=0
\label{eq:app:<H>=U:b}
\end{equation}
For $T=0$ it is:
\begin{equation}
I_\lambda(U,0)=\int_0^{U} de \Omega_\lambda(e) (e-U)(U-e)^{C-1}
\end{equation}
Since $\Omega_\lambda(e) \geq 0$, and $e-U \leq 0$ in the integration domain, we have 
\begin{equation}
I_\lambda(U,0) \leq 0
\end{equation} 
On the other hand for $T\gg U/C$, we find
\begin{equation}
I_\lambda(U,T) \simeq \int_0^{CT} de \Omega_\lambda(e) (e-U)(CT-e)^{C-1}
\end{equation}
where we neglected the terms $U$ as compared to $CT$. By making the change of variable $x=CT-e$, and neglecting again the term $U$ as compared to $CT$, we obtain:
\begin{equation}
I_\lambda(U,T) \simeq \int_0^{CT} dx \Omega_\lambda(CT-x) (CT-x)x^{C-1}
\end{equation}
All three terms forming the integrand are nonnegative, hence:
\begin{equation}
I_\lambda(U,T\gg U/C) \geq 0
\end{equation}
Thus $I_\lambda(U,T)$ is nonpositive for $T=0$ and nonnegative for very large $T$. This implies, that there must be at least one \emph{nonnegative} value of $T$, for which  $I_\lambda(U,T)=0$. Uniqueness, however is not guaranteed.

In a similar way it is also possible to prove that 
\begin{equation}
I_\lambda(0,T)\geq 0\phantom{x}, \qquad I_\lambda(U \gg CT,T)\leq 0
\end{equation}
showing that one can also fix $T$ and find a $U$ such that $I_\lambda(U,T) =0$. Also in this case only existence is guaranteed but not uniqueness. 

Examples for which two or more different energies correspond to the same temperature were reported in \cite{DunkelPRE06,DunkelPHYSA06} for microcanonical ($C=0$) gases with inter-particle interaction of the Lennard-Jones type. These systems undergo a microcanonical phase transition whose signature is the appearance of oscillations in the function $T(U)$, which, therefore, is not invertible (i.e, $U(T)$ is multivalued). These oscillations are expected to appear also if these Lennard-Jones type systems are thermalized by means of a finite bath with specific heat $C>0$. Based on the observation that no oscillation appear in the canonical treatment \cite{DunkelPHYSA06}, one expects that the amplitude of these oscillations decreases with increasing $C$.

\end{document}